\documentclass[12pt]{article}
\usepackage{amsmath,amssymb,amsthm}

\theoremstyle{definition}

\theoremstyle{remark}

\newcommand{\beq}{\begin{eqnarray}}
\newcommand{\eeq}{\end{eqnarray}}
\newcommand{\beqnn}{\begin{eqnarray*}}
\newcommand{\eeqnn}{\end{eqnarray*}}

\newcommand{\tp}[1]{\:{}^{\mathrm{t}}#1}
\newcommand{\CC}{\mathbf{C}}
\newcommand{\PP}{\mathbf{P}}

\newcommand{\ZZ}{\mathbf{Z}}
\newcommand{\bst}{\boldsymbol{t}}

\newcommand{\bszero}{\boldsymbol{0}}
\newcommand{\calO}{\mathcal{O}}
\newcommand{\calP}{\mathcal{P}}

\begin{document}

\title{Integrable structure of\\
modified melting crystal model}
\author{Kanehisa Takasaki\thanks{takasaki@math.h.kyoto-u.ac.jp}\\
{\normalsize Graduate School of Human and Environmental Studies, 
Kyoto University}\\
{\normalsize Yoshida, Sakyo, Kyoto 606-8501, Japan} }
\date{\normalsize IGST12, Zurich, August 20--24, 2012}
\maketitle

\begin{abstract}
Our previous work on a hidden integrable structure of 
the melting crystal model (the $U(1)$ Nekrasov function) 
is extended to a modified crystal model.  
As in the previous case, ``shift symmetries'' 
of a quantum torus algebra plays a central role. 
With the aid of these algebraic relations, 
the partition function of the modified model is shown 
to be a tau function of the 2D Toda hierarchy. 
We conjecture that this tau function belongs to a class 
of solutions (the so called Toeplitz reduction) 
related to the Ablowitz-Ladik hierarchy.  
\end{abstract}

\section{Introduction}

In a previous paper \cite{NT07}, we studied 
a hidden integrable structure in the melting crystal model 
(equivalently, the $U(1)$ Nekrasov function). 
Deforming the model by a charge variable $s$ 
and external potentials with coupling constants 
$\bst = (t_1,t_2,\ldots)$, we showed that 
the partition function coincides, up to simple factors, 
with a tau function of the 1D Toda hierarchy.  
A technical clue is a set of special algebraic relations 
(referred to as ``shift symmetries'') 
among the basis of a quantum torus algebra. 
With the aid of these relations, we could rewrite 
the partition function to a product of simple factors 
and the 1D Toda tau function.  

In this report, we present similar results 
for a modified melting crystal model.  
In the context of topological string theory, 
this model is related to the resolved conifold, 
or local $\CC\PP^1$ geometry of the type 
$\calO(-1)\oplus\calO(-1)\to\CC\PP^1$, 
whereas the previous model corresponds 
to local $\CC\PP^1$ geometry of the type 
$\calO\oplus\calO(-2)\to\CC\PP^1$.  
As it turns out, a hidden integrable structure 
is a non-1D reduction of the 2D Toda hierarchy.  
We conjecture that this reduction will be 
the so called Toeplitz reduction \cite{BCR11}, 
hence the relevant integrable hierarchy 
will be the Ablowitz-Ladik hierarchy.

\section{Modified melting crystal model}

\subsection{Fermions}

We use the same formulation of fermions as our previous work \cite{NT07}:
\begin{itemize}
\item Fourier modes of 
2D complex fermion fields 
\beqnn
  \psi(z) = \sum_{n\in\ZZ}\psi_nz^{-n-1},\quad
  \psi^*(z) = \sum_{n\in\ZZ}\psi^*_nz^{-n}. 
\eeqnn
with anti-commutation relations
\beqnn
  \psi_m\psi^*_n + \psi^*_m\psi_n = \delta_{m+n,0},\quad 
  \psi_m\psi_n + \psi_m\psi_n 
  = \psi^*_m\psi^*_n + \psi^*_n\psi^*_m = 0.
\eeqnn
\item Ground and exited states 
\beqnn
\begin{gathered}
  \langle s| = \langle-\infty|\cdots\psi^*_{s-1}\psi^*_s, \quad 
  |s\rangle = \psi_{-s}\psi_{-s+1}\cdots|-\infty\rangle,\\
  \langle\mu,s| = \langle -\infty|\cdots\psi^*_{\mu_2+s-1}\psi^*_{\mu_1+s},\quad
  |\mu,s\rangle = \psi_{-\mu_1-s}\psi_{-\mu_2-s+1}\cdots|-\infty\rangle.
\end{gathered}
\eeqnn
in the charge-$s$.  The excited states are labelled 
by the set $\calP$ of all partitions $\mu = (\mu_i)_{i=1}^\infty$, 
$\mu_1 \ge \mu_2 \ge \cdots \ge 0$, of arbitrary lengths.\\
\item Special fermion bilinears 
\beqnn
  J_k = \sum_{n\in\ZZ}{:}\psi_{k-n}\psi^*_n{:},\quad
  L_0 = \sum_{n\in\ZZ}n{:}\psi_{-n}\psi^*_n{:},\quad
  W_0 = \sum_{n\in\ZZ}n^2{:}\psi_{-n}\psi^*_n{:}. 
\eeqnn
$J_0$, $L_0$ and $W_0$ are zero-modes of $U(1)$ current, 
Virasoro and $W^{(3)}$ algebras. 
\end{itemize}

\subsection{Partition function in fermionic form}

Our previous melting crystal model \cite{NT07} is defined by 
the partition function 
\beq
  Z(s,\bst) = \langle s|G_{+}q^{lW_0/2}Q^{L_0}e^{H(\bst)}G_{-}|s\rangle, 
\eeq
where 
\begin{itemize}
\item  $q$ and $Q$ are constant in the range 
$0 < |q| < 1$ and $0 < |Q| < 1$, and $l$ is an integer.  
\item  $H(\bst)$ is a linear combination 
$\displaystyle H(\bst) = \sum_{k=1}^\infty t_kH_k$ of 
the special fermion bilinears  
\beqnn
  H_k = \sum_{n\in\ZZ}q^{kn}{:}\psi_{-n}\psi^*_n{:}, 
  \quad k \in \ZZ. 
\eeqnn
\item  $G_{\pm}$ are the transfer operators 
\beqnn
  G_{\pm} = \exp\left(\sum_{k=1}^\infty
            \frac{q^{k/2}}{k(1-q^k)}J_{\pm k}\right) 
\eeqnn
of Okounkov and Reshetikhin \cite{ORV03}.  
\end{itemize}
We now modify this model as follows: 
\begin{itemize}
\item[(i)]Replace $H(\bst)$ with $\displaystyle 
H(\bst,\hat{\bst}) = \sum_{k=1}^\infty t_kH_k 
+ \sum_{k=1}^\infty\hat{t}_kH_{-k}$. 
\item[(ii)]Replace $G_{-}$ with one of another pair 
of Okounkov and Pandharipande's transfer operators 
\beqnn
  G'_{\pm} = \exp\left(- \sum_{k=1}^\infty 
             \frac{(-1)^kq^{k/2}}{k(1-q^k)}J_{\pm k}\right).
\eeqnn
\end{itemize}
The partition function of the modified model is thereby defined as 
\beq
  Z'(s,\bst,\hat{\bst}) 
  = \langle s|G_{+}q^{lW_0/2}Q^{L_0}e^{H(\bst,\bar{\bst})}G'_{-}|s\rangle.
\eeq

\subsection{Partition function as sum over partitions}

According to Okounkov, Reshetikhin and Vafa \cite{ORV03}, 
$G_{+}$ and $G'_{-}$ act on the ground states 
$\langle s|$, $|s\rangle$ as 
\beq
  \langle s|G_{+} = \sum_{\mu\in\calP}\langle\mu,s|s_\mu(q^{-\rho}),\quad
  G'_{-}|s\rangle = \sum_{\mu\in\calP}s_{\tp{\mu}}(q^{-\rho})|\mu,s\rangle, 
\eeq
where $s_\mu(q^{-\rho})$'s are special values of the Schur functions at 
\beqnn
  q^{-\rho} = (q^{1/2},q^{3/2},\ldots,q^{n-1/2},\dots).
\eeqnn
These special values are known to have the hook length formula 
\beq
  s_\mu(q^{-\rho}) 
  = \frac{q^{-\kappa(\mu)/4}}
    {\prod_{(i,j)\in\mu}(q^{-h(i,j)/2}-q^{h(i,j)/2})}, 
\eeq
where $h(i,j)$ denotes the length of the hook with corner 
at the cell $(i,j)$ of the Young diagram. 
$\tp{\mu}$ denotes the conjugate (or transpose) of $\mu$. 
Since the operators $q^{lW_0/2}$, $Q^{L_0}$ and 
$e^{H(\bst,\hat{\bst})}$ are diagonal with respect 
to $\langle\mu,s|$'s and $|\mu,s\rangle$'s, 
$Z'(s,\bst,\bar{\bst})$ can be expanded 
to a single sum over $\calP$ as 
\beq
  Z'(s,\bst,\hat{\bst}) 
  = \sum_{\mu\in\calP}
     s_\mu(q^{-\rho})s_{\tp{\mu}}(q^{-\rho}) 
     q^{l\langle\mu,s|W_0|\mu,s\rangle/2}
     Q^{\langle\mu,s|L_0|\mu,s\rangle}
     e^{\langle\mu,s|H(\bst,\bar{\bst})|\mu,s\rangle}. 
\eeq
The diagonal matrix elements of $L_0$ and $W_o$ can be expressed as 
\beq
\begin{gathered}
  \langle\mu,s|L_0|\mu,s\rangle = |\mu| + \frac{s(s+1)}{2},\\
  \langle\mu,s|W_0|\mu,s\rangle
    = \kappa(\mu) + (2s+1)|\mu| + \frac{s(s+1)(2s+1)}{6}, 
\end{gathered}
\eeq
where 
\beqnn
  \displaystyle |\mu| = \sum_{i\ge 1}\mu_i,\quad 
  \kappa(\mu) = \sum_{i\ge 1}\mu_i(\mu_i - 2i + 1). 
\eeqnn
The diagonal matrix elements of $H(\bst,\hat{\bst})$ 
give potentials of the form 
\beq
\begin{gathered}
  \langle\mu,s|H(\bst,\bar{\bst})|\mu,s\rangle 
  = \sum_{k=1}^\infty t_k\Phi_k(\mu,s) 
    + \sum_{k=1}^\infty\hat{t}_k\Phi_{-k}(\mu,s),\\
  \Phi_k(\mu,s) 
  = \sum_{i=1}^\infty(q^{k(s+\mu_i-i+1)} - q^{k(s-i+1)}) 
    + \frac{q^k(1-q^{ks})}{1-q^k}. 
\end{gathered}
\eeq

When $\bst = \hat{\bst} = \bszero$ and $s = 0$, 
this partition function reduces to 
\beq
  Z' = \sum_{\mu\in\calP}s_\mu(q^{-\rho})s_{\tp{\mu}}(q^{-\rho}) 
        q^{l\kappa(\mu)/2}(q^{l/2}Q)^{|\mu|}. 
\eeq
Up to a sign factor, this coincides with the sum 
derived by Brian and Pandharipande \cite{BP08} from 
the Gromov-Witten theory of local curves. 
This sum was further studied by Caporaso et al \cite{CGMPS06} 
in the context of toric topological string theory. 
We need the coupling constants $\bst,\hat{\bst}$ 
and the charge variable $s$ to formulate an integrable structure.

\section{Hidden integrable structure}

\subsection{Shift symmetries in quantum torus algebra}

The fermion bilinears 
\beqnn
  V^{(k)}_m = q^{-km/2}\sum_{n\in\ZZ}q^{kn}{:}\psi_{n-m}\psi^*_n{:},\quad 
  k,m \in \ZZ, 
\eeqnn
satisfy the commutation relations 
\beq
  [V^{(k)}_m,V^{(l)}_n] = (q^{(lm-kn)/2} - q^{(kn-lm)/2}) 
  \left(V^{(k+l)}_{m+n} - \delta_{m+n,0}\frac{q^{k+l}}{1-q^{k+l}}\right) 
\eeq
of (a central extension of) the quantum torus algebra.  
Note that the c-number term on the right hand side 
turns into $-m\delta_{m+n}$ as $k + l \to 0$.  

The following algebraic relations, referred to as 
{\it shift symmetries\/}, play a central role 
in identifying a hidden integrable structure of the partition function: 

\paragraph*{(i) First Symmetries}
\begin{gather}
  G_{-}G_{+}\left(V^{(k)}_m - \delta_{m,0}\frac{q^k}{1-q^k}
    \right)(G_{-}G_{+})^{-1} 
  = (-1)^k\left(V^{(k)}_{m+k} - \delta_{m+k,0}\frac{q^k}{1-q^k}\right), \\
  G'_{-}G'_{+}\left(V^{(-k)}_m - \delta_{m,0}\frac{1}{1-q^k}
    \right)(G'_{-}G'_{+})^{-1} 
  = V^{(-k)}_{m+k} - \delta_{m+k,0}\frac{1}{1-q^k} 
\end{gather}
for $k > 0$, $m \in \ZZ$.
\paragraph*{(ii) Second symmetries}
\beq
  q^{W_0/2}V^{(k)}_mq^{-W_0/2} = V^{(k-m)}_m 
\eeq
for $k,m \in \ZZ$.

\subsection{Rewriting partition function to tau function}

With the aid of the shift symmetries, we can rewrite 
the partition function to a {\it tau function\/}. 

\paragraph{First step:} 
Since $H_k = V^{(k)}_0$ and $J_k = V^{(0)}_k$, 
the shift symmetries with respect to $G_{\pm}$'s and $q^{W_0/2}$ 
yields the relation 
\beqnn
  G_{+}H_kG_{+}^{-1} 
  = (-1)^kG_{-}^{-1}q^{-W_0/2}J_kq^{W_0/2}G_{-} 
    + \frac{q^k}{1-q^k}. 
\eeqnn
This implies that 
\begin{multline*}
  G_{+}\exp\left(\sum_{k=1}^\infty t_kH_k\right)G_{+}^{-1}
  = \exp\left(\sum_{k=1}^\infty\frac{q^kt_k}{1-q^k}\right)\\
  \mbox{}\times 
    G_{-}^{-1}q^{-W_0/2}\exp\left(
      \sum_{k=1}^\infty(-1)^kt_kJ_k\right)q^{W_0/2}G_{-}.
\end{multline*}
\paragraph{Second step:}
In the same way, by the shift symmetries with respect 
to $G'_{\pm}$'s and $q^{W_0/2}$, we have the relation 
\beqnn
  {G'_{-}}^{-1}H_{-k}G'_{-} 
  = G'_{+}q^{-W_0/2}J_{-k}q^{W_0/2}{G'_{+}}^{-1} 
  + \frac{1}{1-q^k}, 
\eeqnn
hence 
\begin{multline*}
  {G'_{-}}^{-1}\exp\left(\sum_{k=1}^\infty\hat{t}_kH_{-k}\right)G'_{-} 
  = \exp\left(\sum_{k=1}^\infty\frac{\hat{t}_k}{1-q^k}\right)\\
  \mbox{}\times 
    G'_{+}q^{-W_0/2}\exp\left(
      \sum_{k=1}^\infty\hat{t}_kJ_{-k}\right)
    q^{W_0/2}{G'_{+}}^{-1}. 
\end{multline*}
\paragraph{Third step:}
The foregoing calculations show that the operator 
in the definition of $Z'(s,\bst,\hat{\bst})$ 
can be thus expressed as 
\begin{multline*}
  G_{+}q^{lW_0/2}Q^{L_0}e^{H(\bst,\hat{\bst})}G'_{-}\\
  = \exp\left(\sum_{k=1}^\infty\frac{q^kt_k + \hat{t}_k}{1-q^k}\right)
    G_{-}^{-1}q^{-W_0/2}\exp\left(\sum_{k=1}^\infty(-1)^kt_kJ_k\right)\\
  \mbox{}\times 
    q^{W_0/2}G_{-}G_{+}q^{lW_0/2}Q^{L_0}G'_{-}G'_{+}q^{-W_0/2}
    \exp\left(\sum_{k=1}^\infty\hat{t}_kJ_{-k}\right)
    q^{W_0/2}{G'_{+}}^{-1}. 
\end{multline*}
The leftmost and rightmost operators in this expression 
act on $\langle s|$ and $|s\rangle$ as 
\beqnn
\begin{gathered}
  \langle s|G_{-}^{-1}q^{-W_0/2} 
  = q^{-s(s+1)(2s+1)/12}\langle s|, \\
  q^{W_0/2}{G'_{+}}^{-1}|s\rangle 
  = q^{s(s+1)(2s+1)/12}|s\rangle. 
\end{gathered}
\eeqnn

We thus find the following expression of the partition function: 
\begin{gather}
  Z'(s,\bst,\hat{\bst}) 
  = \exp\left(\sum_{k=1}^\infty\frac{q^kt_k + \hat{t}_k}{1-q^k}\right)
    \tau'(s,-t_1,t_2,-t_3,\ldots,-\hat{t}_1,-\hat{t}_2,-\hat{t}_3,\ldots),\\
  \tau'(s,\bst,\hat{\bst}) 
  = \langle s|\exp\left(\sum_{k=1}^\infty t_kJ_k\right)
    g'\exp\left(- \sum_{k=1}^\infty\hat{t}_kJ_{-k}\right)
    |s\rangle, 
\end{gather}
where 
\beq
  g' = q^{W_0/2}G_{-}G_{+}q^{lW_0/2}Q^{L_0}G'_{-}G'_{+}q^{-W_0/2}. 
\eeq
$\tau'(s,\bst,\hat{\bst})$ is a tau function of 
the {\it 2D Toda hierarchy\/}, in which 
$t_k$'s and $\hat{t}_k$'s are two independent sets 
of time variables. 
It is well known that a tau functions remains 
to be a tau function after multiplied by 
an exponential function of a linear function 
of the time variables.  Hence $Z'(s,\bst,\hat{\bst})$ 
itself is also a tau function.

\subsection{Comparison with previous model}

The partition function 
\beqnn
  Z(s,\bst) = \langle s|G_{+}q^{lW_0/2}Q^{L_0}e^{H(\bst)}G_{-}|s\rangle 
\eeqnn
of the previous model \cite{NT07} can be rewritten as 
\beq
  Z(s,\bst) 
  = \exp\left(\sum_{k=1}^\infty\frac{t_kq^k}{1-q^k}\right)
    q^{-s(s+1)(2s+1)/6}\tau(s,-t_1,t_2,-t_3,\ldots), 
\eeq
where $\tau(s,\bst)$ is a tau function of 
the {\it 1D Toda hierarchy\/}.   
The 1D Toda hierarchy is a special case 
({\it reduction} in the terminology of integrable systems) 
of the 2D Toda hierarchy in which the tau function 
depends on the two sets of time variables $\bst,\hat{\bst}$ 
through their difference as 
\beq
  \tau(s,\bst,\hat{\bst}) = \tau(s,\bst-\hat{\bst}). 
\eeq
The reduced function $\tau(s,\bst)$ becomes 
the tau function of the 1D Toda hierarchy.  
In fact, $\tau(s,\bst)$ can have 
different expressions such as 
\beq
\begin{aligned}
  \tau(s,\bst) 
  &= \langle s|\exp\left(\sum_{k=1}^\infty t_kJ_k\right)g|s \rangle\\
  &= \langle s|\exp\left(\sum_{k=1}^\infty\frac{t_k}{2}J_k\right) 
     g\exp\left(\sum_{k=1}^\infty\frac{t_k}{2}J_{-k}\right) 
     |s \rangle\\
  &= \langle s|g\exp\left(\sum_{k=1}^\infty t_kJ_{-k}\right)|s \rangle, 
\end{aligned}
\eeq
where 
\beq
  g = q^{W_0/2}G_{-}G_{+}q^{lW_0/2}Q^{L_0}G_{-}G_{+}q^{W_0/2}. 
\eeq
This is a consequence of the intertwining relations 
\beq
  J_kg = gJ_{-k} \quad\text{for $k = 1,2,\ldots$} 
\eeq
that can be derived from shift symmetries. 

The tau function $\tau'(s,\bst,\hat{\bst})$ 
of the modified model does not have this property, 
hence a tau function of the 2D Toda hierarchy in a genuine sense.  
Its status in the 2D Toda hierarchy, however, is still obscure.

\subsection{Toeplitz reduction}

In search for the status of $\tau'(s,\bst,\hat{\bst})$, 
let us draw attention to the following {\it FAKE} 
intertwining relations 
\beq
  J_kg' = g'J_k \quad\text{for $k = \pm 1,\pm 2,\ldots$} 
\eeq
referred to as the {\it Toeplitz condition}
in the literature of integrable systems.  
Though details are omitted, we can {\it derive} these relations 
from shift symmetries.  Actually, these relations 
shoud {\it NOT} hold. 
If these relations were correct, we could move $J_{\pm}$'s 
in the tau function to the other side of 
the ground state expectation value and find that 
\beq
  \tau'(s,\bst,\hat{\bst}) 
  = \exp\left(\sum_{k=1}^\infty kt_k\hat{t}_k\right)
    \langle s|g'|s\rangle. 
\eeq
Namely, the tau function would turn out to be 
an almost trivial one.  This is not the case.  

This contradictory situation stems from 
potential inconsistency of shift symmetries.  
Careless use of shift symmetries can lead to wrong results.  
This inconsistency seems to be related 
to non-associativity of some operators 
on the fermionic Fock space.  We have been unable 
to establish a fully consistent theory of shift symmetries. 

In spite of apparent inconsistency, we are inspired 
by the fake intertwining relations to conjecture 
that the tau function $\tau'(s,\bst,\hat{\bst})$ 
belongs to the {\it Toeplitz reduction\/} 
of the 2D Toda hierarchy \cite{BCR11}, hence a solution 
of the {\it Ablowitz-Ladik hierarchy\/}. 
This conjecture is also partly supported by 
the work of Brini on the resolved conifold \cite{Brini10}.  
Note that our partition function in the case of $l = 0$ 
is related to the partition function of topological strings 
on the resolved conifold.  Another possible test 
towards this conjecture is to examine the thermodynamic limit 
as we have done for the previous model \cite{NT11}.

\subsection*{Acknowledgements}

This work is partly supported by JSPS Grants-in-Aid 
for Scientific Research No. 22540186 and No. 24540223 
from the Japan Society for the Promotion of Science.


\begin{thebibliography}{99}

\bibitem{NT07}
T.~Nakatsu and K.~Takasaki, 
Melting crystal, quantum torus and Toda hierarchy, 
Comm. Math. Phys. {\bf 285} (2009), 445--468 
(arXiv:0710.5339 [hep-th]). 


\bibitem{BCR11}
A.~Brini, G.~Carlet and P.~Rossi, 
Integrable hierarchies and the mirror model of local $\mathbb{CP}^1$, 
arXiv:1105.4508 [math.AG]. 

\bibitem{ORV03}
A.~Okounkov, N.~Reshetikhin and C.~Vafa,  
Quantum Calabi-Yau and classical crystals, 
in: P. Etingof, V. Retakh and I.M. Singer (eds.), 
{\it The unity of mathematics\/}, 
Progr. Math.  vol. 244 (Birkh\"auser, 2006), pp. 597--618 
(arXiv:hep-th/0309208). 

\bibitem{BP08}
J.~Bryan and R.~Pandharipande, 
The local Gromov-Witten theory of curves, 
J. Amer. Math. Soc. {\bf 21} (2008), 101--136 
(arXiv:math/0411037). 

\bibitem{CGMPS06}
N.~Caporaso, L.~Griguolo, M.~Mari\~{n}o, S.~Pasquetti and D.~Seminara, 
Phase transitions, double-scaling limit, and topological strings, 
Phys. Rev. {\bf D75} (2007), 046004 
(arXiv:hep-th/0606120). 

\bibitem{Brini10}
A.~Brini, The local Gromov-Witten theory of 
$\mathbb{CP}^1$ and integrable hierarchies, 
arXiv:1002.0582 [math-ph]. 

\bibitem{NT11}
T.~Nakatsu and K.~Takasaki, 
Thermodynamic limit of random partitions 
and dispersionless Toda hierarchy, 
J. Phys. A: Math. Theor. {\bf 45} (2012), 025403 (38pp) 
(arXiv:1110.0657 [math-ph]). 





\end{thebibliography}
\end{document}